\documentclass[]{spie}  

\usepackage{amsmath,amsfonts,amssymb}
\usepackage{graphicx}
\usepackage[colorlinks=true, allcolors=blue]{hyperref}

\title{The Monitoring Logging and Alarm System of the ASTRI Mini-Array gamma-ray air-Cherenkov experiment at the Observatorio del Teide}

\author{Federico Incardona$^\text{a}$, Alessandro Costa$^\text{a}$, Kevin Munari$^\text{a}$, Salvatore Gambadoro$^\text{a,e}$, Stefano Germani$^\text{b}$, Pietro Bruno$^\text{a}$, Andrea Bulgarelli$^\text{d}$, Vito Conforti$^\text{d}$, Fulvio Gianotti$^\text{d}$, Alessandro Grillo$^\text{a}$, Valerio Pastore$^\text{d}$, Federico Russo$^\text{d}$, Joseph Schwarz$^\text{c}$, Gino Tosti$^\text{b}$, and Salvatore Cavalieri$^\text{e}$, for the ASTRI Project}

\affil[\space]{$^\text{a}$INAF, Osservatorio Astrofisico di Catania, Via S Sofia 78, I-95123 Catania, ITALY}
\affil[\space]{$^\text{b}$Universit\`a di Perugia, Dipartimento di Fisica e Geologia, IT}
\affil[\space]{$^\text{c}$INAF, Osservatorio Astronomico di Brera, IT}
\affil[\space]{$^\text{d}$INAF, Osservatorio di Astrofisica e Scienza dello Spazio di Bologna, IT}
\affil[\space]{$^\text{e}$Universit\`a degli Studi di Catania, Dipartimento Ingegneria Elettrica Elettronica Informatica, IT}
\affil[\space]{$^1$\url{http://www.astri.inaf.it/en/library/}}

\authorinfo{Further author information: (Send correspondence to F. Incardona)\\F. Incardona: E-mail: federico.incardona@inaf.it}

\begin{document} 
\maketitle

\begin{abstract}
The ASTRI Mini-Array is a project for the Cherenkov astronomy in the TeV energy range. ASTRI Mini-Array consists of nine Imaging Atmospheric Cherenkov telescopes located at the Teide Observatory (Canarias Islands). 
Large volumes of monitoring and logging data result from the operation of a large-scale astrophysical observatory. In the last few years, several “Big Data” technologies have been developed to deal with such volumes of data, especially in the Internet of Things (IoT) framework.
We present the Monitoring, Logging, and Alarm (MLA) system for the ASTRI Mini-Array aimed at supporting the analysis of scientific data and improving the operational activities of the telescope facility. The MLA system was designed and built considering the latest software tools and concepts coming from Big Data and IoT to respond to the challenges posed by the operation of the array. A particular relevance has been given to satisfying the reliability, availability, and maintainability requirements towards all the array sub-systems and auxiliary devices.
The system architecture has been designed to scale up with the number of devices to be monitored and with the number of software components to be considered in the distributed logging system.

\end{abstract}


\section{Introduction}

The ASTRI (\textit{Astrofisica con Specchi a Tecnologia Replicante Italiana}) project started in 2010 to support the development of technologies within the Cherenkov Telescope Array \citenum{acharya2013introducing}. The ASTRI  Mini-Array project \citenum{pareschi2016astri} has expanded becoming an international effort led by the Italian National Institute for Astrophysics (INAF). It is planned to consist of nine Cherenkov telescopes and cameras in full operations \citenum{SCUDERI2022}, \citenum{VERCELLONE2022}, and is based on the experience provided by the ASTRI prototype. \\
Internet of Things (IoT) is an emerging technology that is becoming a major topic of interest among technology giants and business communities. The data generated by IoT devices is large in volume and needs to be analyzed using Big Data analytics engine (see, e.g. \citenum{pena2018framework}) in order to extract the critical information and detect behavioral patterns. \\
This paper presents the monitoring, logging, and alarm software architecture currently under development for the ASTRI Mini-Array (see, also \citenum{Costa:2021c8}). This architecture takes advantage of continuing technological evolution \citenum{costa:icalepcs2019-mopha032} to respond to the challenges posed by the operation of the array, in particular, to satisfy the reliability, availability, and maintainability requirements of all its sub-systems and auxiliary devices. The system architecture is based on the ALMA Common Software\footnote{ACS: \url{https://www.eso.org/projects/alma/develop/acs/}} (ACS \citenum{chiozzi2004alma}) and has been designed to scale up with the number of devices to be monitored and with the number of software components to be taken into account in the distributed logging system.      

\section{Logging, Monitoring and Alarm Systems Architecture}

The Monitoring, Logging, and Alarm (MLA) System monitors the overall performance of the ASTRI Mini-Array through the acquisition of environmental housekeeping data, log files, and alarms from instruments, and generates status reports or notifications to the operator via an Operator Human Machine Interface (HMI). In the context of ASTRI Software \citenum{BulgarelliSPIE}, the HMI provides the interested users with a set of different Graphical User Interfaces (GUIs), accessible from the web, desktop, and mobile clients, through which they can watch in real-time the data streaming, and can browse and query the historical one. It is made up of a backend exposing data through HTTP/GraphQL\footnote{HTTP/GraphQL: \url{https://graphql.org/learn/serving-over-http/}} APIs and Websockets\footnote{The WebSocket Protocol: \url{https://tools.ietf.org/html/rfc6455}} to multiple front-ends. \\
The Alarm System provides the service that gathers, filters, exposes, and persists all the relevant alarms raised by both devices (such as telescopes and other devices) and Supervisory Control And Data Acquisition (SCADA) processes. The monitoring, logging, and alarm data are saved in the MLA Archives, which are databases optimized for real-time applications \citenum{IncardADASS} (high read-write throughput\footnote{Apache Cassandra: \url{https://cassandra.apache.org/}}) used to keep historical events. These archives are able to resolve any queries coming from operators and engineering GUIs and are periodically transferred to the ASTRI Data Center.

\section{The Monitoring System}

\begin{figure}
    \centering
    \includegraphics[trim={0.5cm 3cm 0.5cm 4cm},clip, width=0.65\textwidth]{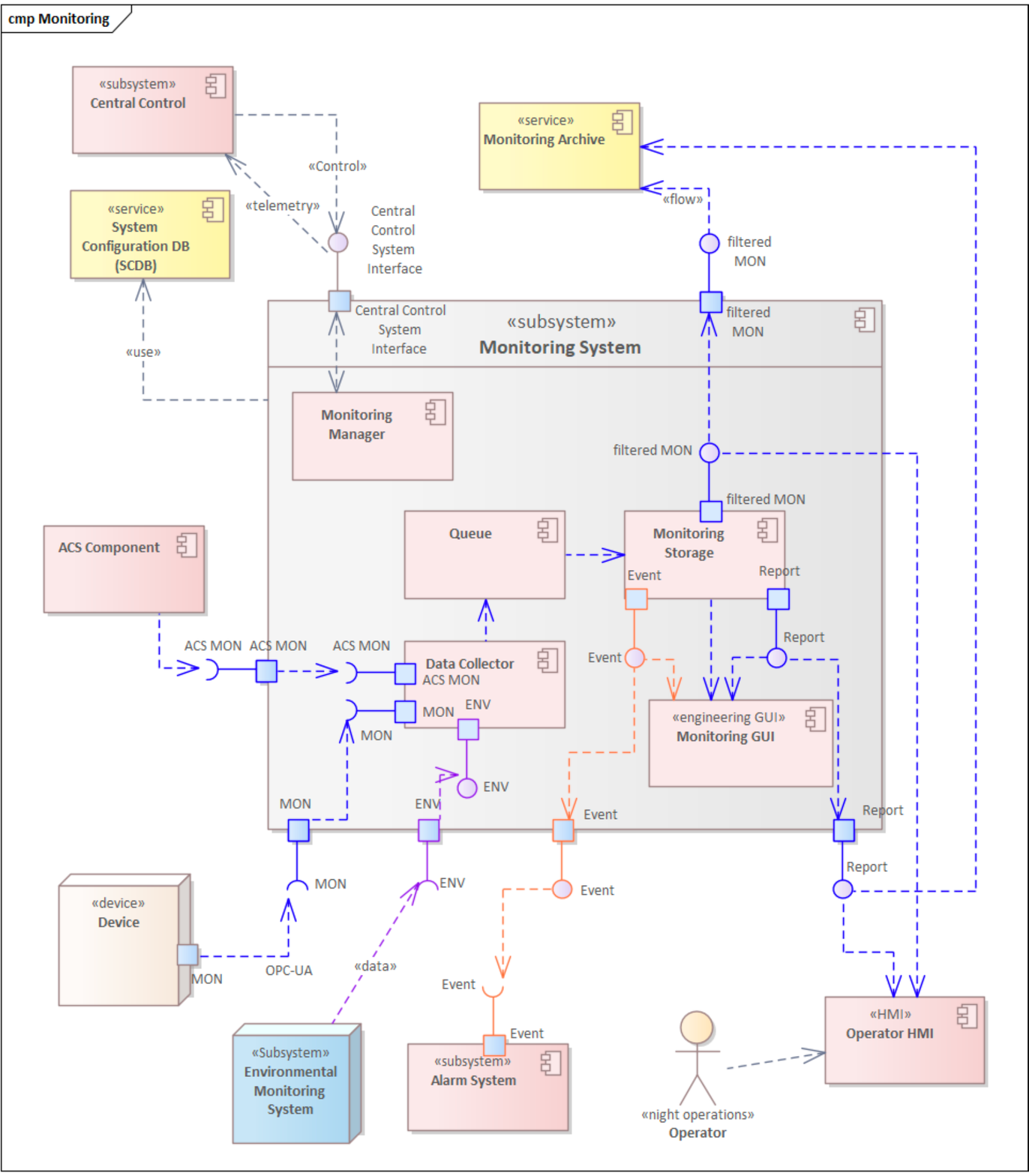}
    \caption{Monitoring System Architecture}
    \label{fig_1}
\end{figure}

The ASTRI Mini-Array system will generate a significant amount of monitoring data, collecting information on the performance of a variety of critical and complex electrical and mechanical components. We foresee about 20000 monitoring points, and the updating frequency will be no higher than about 1 Hz, except for short-term debugging campaigns that may require more frequent updates. This monitoring information enriches and complements observational data, and is crucial for most troubleshooting efforts performed by engineering teams. The Monitoring System (Fig. \ref{fig_1}) will enable the staff to use a systematic approach to fault detection and diagnosis and support corrective and predictive maintenance to minimize the downtime of the system.
\\The Monitoring System provides the services that gather monitoring data from the telescopes, the Environmental Monitoring System, and other instruments, and saves them in the Monitoring Archive. It also provides a framework for the evaluation and analysis of abnormal situations. The Monitoring System works continuously to record any monitoring data. It provides a global base support service and is supposed to be up, running, and available to any client that wants to use it at all times. The Monitoring System functionality can be broadly classified into three main areas: collection, persistence, and (limited) processing. 
\begin{itemize}
    \item Collection: anything collected by the monitoring subsystem is associated with a time-stamp. Monitoring System collects sensor data and other similar data that change over time, status, and other information.
    \item  Persistence: most monitoring data are persisted for later analysis and processing. The persisting may operate on the raw monitoring data, or on slightly processed material. The Monitoring System is able to acquire, process, and save environmental condition data and monitoring points into the Monitoring Archive.
    \item Processing capabilities: (i) suppression of duplicate values, e.g., if there is no need to keep repeating with full frequency a sensor’s value or a component’s state as long as that value or state does not change; (ii) comparison functions, so that processing emits a value only if an incoming monitoring point lies within a predefined window of values; (iii) statistical processing; (iv) sensor values filtering, depending on their status. 
\end{itemize}
The Monitoring System allows also browsing of the monitoring data offline either to reconstruct past events or for further investigations, and it provides the capability to re-sample the monitoring information coming in the form of irregular and unevenly-spaced time-series data to a consistent and regular frequency. 

The breakdown structure of the Monitoring System is the following (see Fig. \ref{fig_1}):
\begin{itemize}
    \item \textbf{Data Collector} is responsible for retrieving monitoring points from the hardware devices (through the OPC-UA\footnote{The OPC Unified Architecture (UA): \url{https://opcfoundation.org/about/opc-technologies/opc-ua/}} and ACS components \citenum{chiozzi2004alma}). The Data Collector is also in charge of data processing, filtering out duplicate values, and normalizing retrieved data in a common format (i.e., Apache AVRO \footnote{Apache AVRO: \url{https://avro.apache.org/}}). It reads the data sources (i.e., OPC-UA and ACS endpoints) and the set of nodes from which collecting data from the Configuration Database (CDB), which is responsible for storing the System Configuration Data Model. The Data Collector is optimized for efficiency. In particular, the OPC-UA subscriptions are grouped by the sampling interval.
    \item \textbf{Mon Queue} takes formatted and validated monitoring points from the Data Collector. It uses a very fast-in-memory database discarding data after a given retention period and serves as a: (i) buffer to synchronize data collectors, (ii) dispatchers that operate at different speeds, (iii) subscription/notification service to any clients interested in receiving streaming of real-time monitoring data.
    \item \textbf{Monitoring Point Dispatcher} consumes data from the Mon Queue, delivering them to a long-term data storage and to the HMI.
    \item \textbf{Monitor Manager} acts as a coordinator: its tasks are to start and stop the entire subsystem and to provide the current status of the monitoring subsystem components.
\end{itemize}

\section{Logging System}
The ASTRI Mini-Array System can generate a significant amount of log files, in the order of about 200 Mbps, and the logging architecture must be designed taking into account its impact on the performance of the system as a whole. Particular attention has been paid to enabling filtering of log events both at the device and central level. Such filtering capability is based on log priority and it can be configured at the system level. Logs will not only be useful to diagnose failures detected during operation but they will also be needed for long-term performance analysis.
\begin{figure}
    \centering
    \includegraphics[trim={1cm 4.5cm 0.2cm 4.5cm},clip,angle=-90,width=0.8\textwidth]{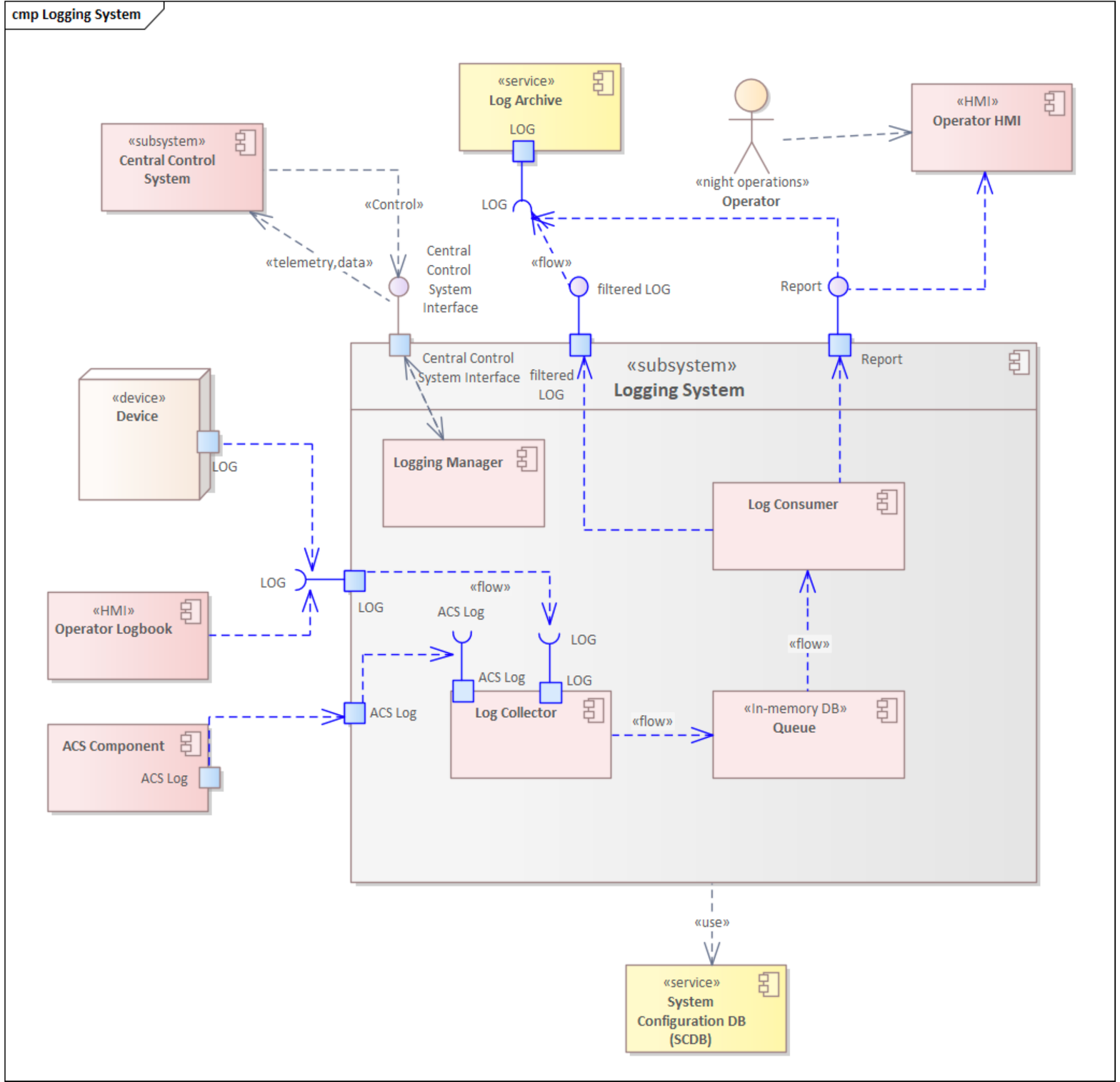}
    \vspace{10pt}\caption{Logging System Architecture}
    \label{fig_2}
\end{figure}
\\The Logging System (LOUD \citenum{Incardona:2021WK}) (Fig. \ref{fig_2}) gets logging information from relevant software and hardware components. It ingests software logs produced by: (i) subsystems using the control framework, (ii) observation scripts, and (iii) low-level firmware, which requires reformatting to adapt to the rest of the logs, (iv) hardware systems, (v) records of actions of the user over the HMIs, in order to keep track of the activities that the operators perform on the system.
 
The breakdown structure of the Logging System is the following (see Fig. \ref{fig_2}): 
\begin{itemize}
    \item \textbf{Log Collector} is responsible for reading and collecting log records from the system components and User Interfaces. It also processes them, generating a log entry according to the log data model and sending them to the Log Queue. The list of the component endpoints is stored in the Configuration Database.
    \item \textbf{Log Queue} acts as a buffer to synchronize the other components of the subsystem that could operate at different speeds.
    \item \textbf{Log Consumer} sends log entries to the Log Archive and communicates with the Operator HMI to provide selected near-real-time logging information.
    \item \textbf{Logging Manager} acts as a coordinator: its tasks are to start and stop the entire subsystem and to provide the current status of the logging subsystem components.
\end{itemize}

\section{Alarm System}
\begin{figure}
    \centering
    \includegraphics[trim={1cm 4.5cm 5.5cm 4.5cm},clip,angle=-90,width=0.8\textwidth]{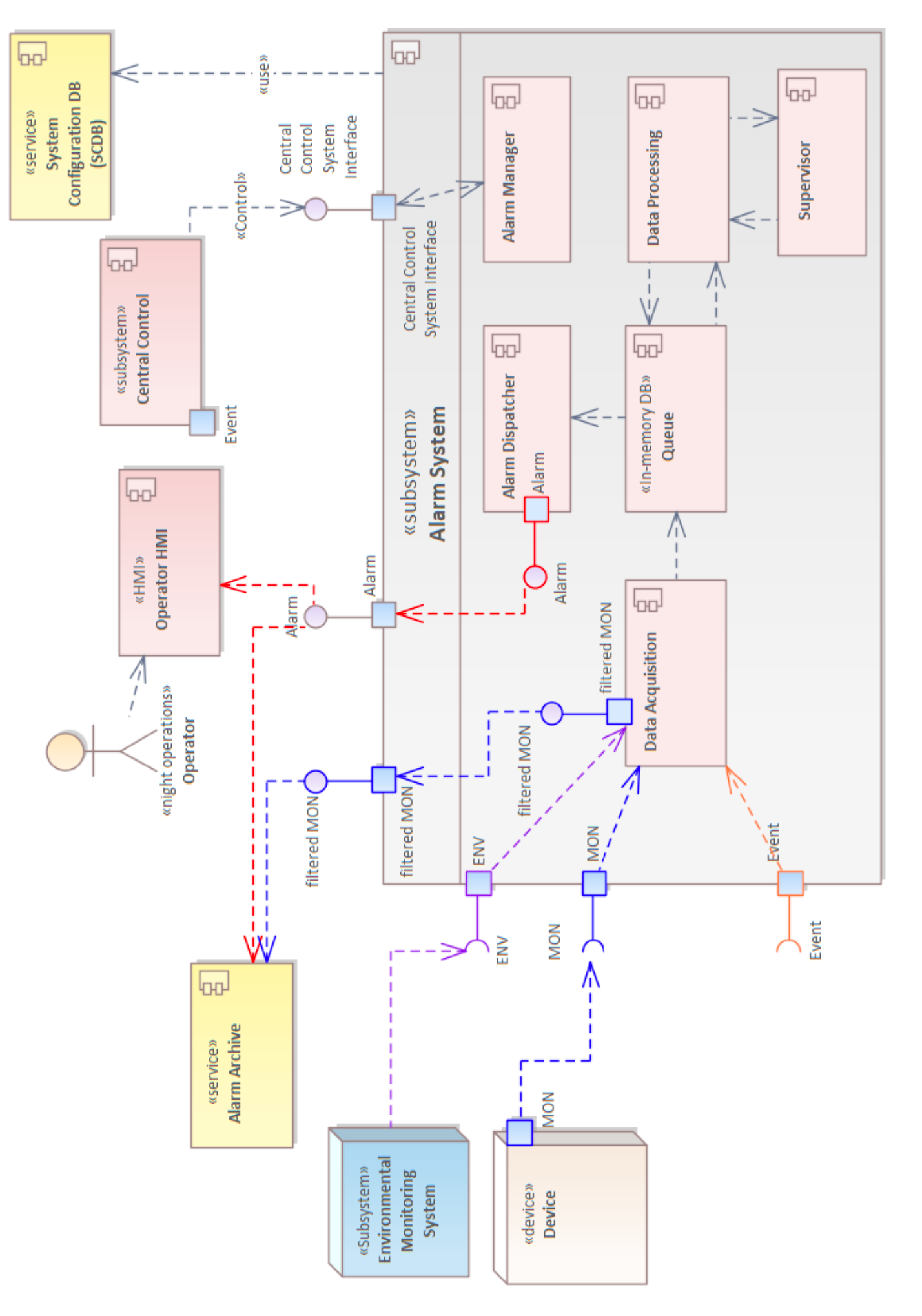}
    \vspace{10pt}\caption{Alarm System Architecture}
    \label{fig_3}
\end{figure}

\begin{figure}
    \centering
    \includegraphics[width=1\textwidth]{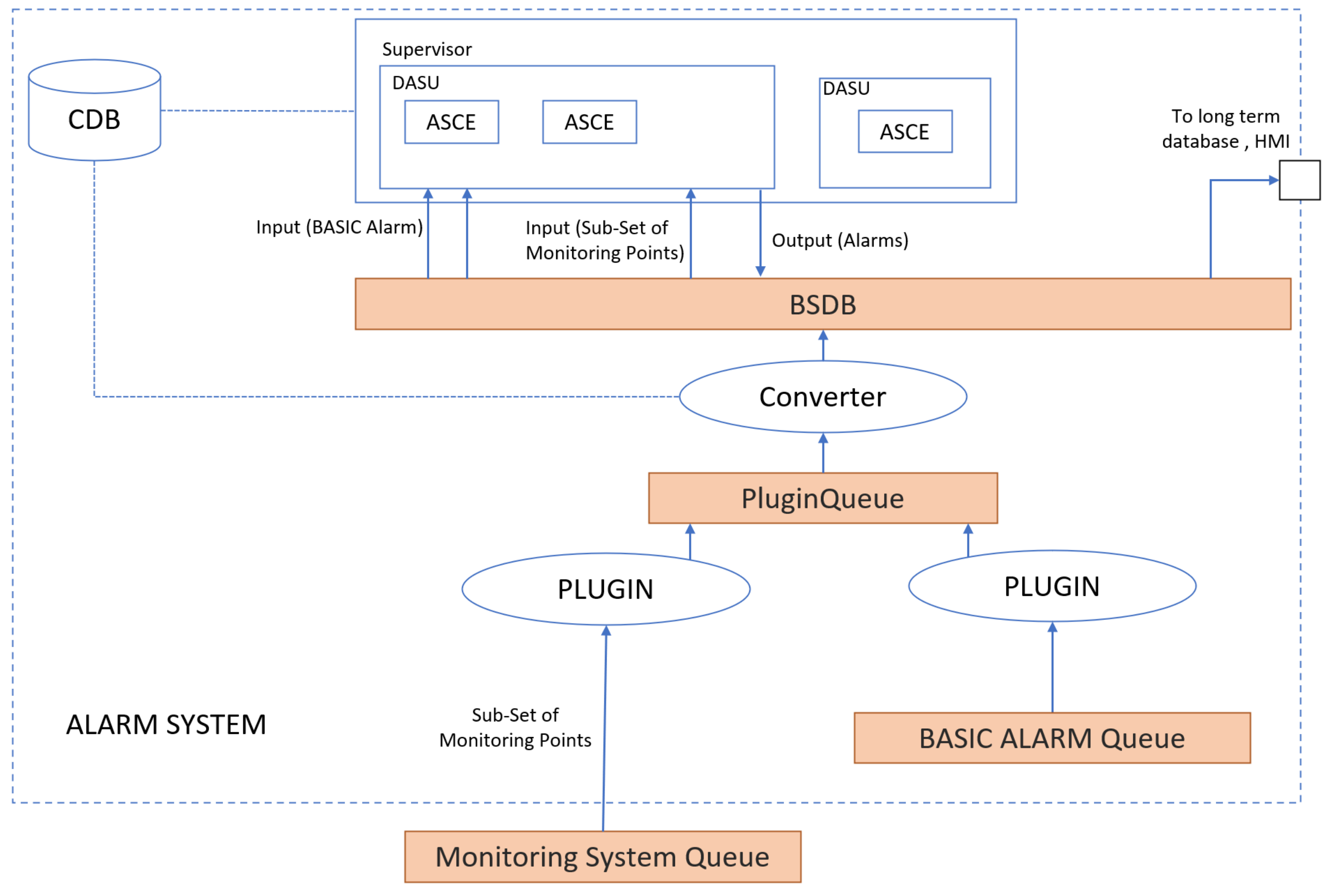}
    \caption{Alarm System Queue}
    \label{fig_4}
\end{figure}

The Alarm System (AS) (Fig. \ref{fig_3}) provides the service that gathers, filters, exposes, and persists all the relevant alarms raised by devices (such as the telescopes) and software processes (e.g., Monitoring System, Logging System, Array Data Acquisition System \citenum{ConfortiSPIE, PastoreSPIE}, Telescope Control System \citenum{RussoSPIE}, etc.). It also creates and filters new alarms based on a selection of the most critical monitoring points. The alarms, which by definition require human attention and response, are sent to Operators via the HMI. The current implementation of the alarm system is a customization of the Integrated Alarm System\footnote{IAS: \url{https://integratedalarmsystem-group.github.io}} (IAS \citenum{IAS}) (Fig. \ref{fig_4}).

The breakdown structure of the Alarm System is the following (see Fig. \ref{fig_3}):
\begin{itemize}
    \item \textbf{Data Acquisition} is a software component that interfaces with the ACS remote control software. It is also in charge of filtering out and discarding duplicate values. The acquisition is performed by the so-called \textit{PluginQueue} (Fig. \ref{fig_4}) that sends the collected Basic Alarm events (from ACS) to a dedicated Apache Kafka \footnote{Apache Kafka \url{https://kafka.apache.org}} data stream.
    \item \textbf{Data Processing} is performed by the Converter in the IAS context. It takes the data points from the PluginQueue and normalizes them in a common data structure, as defined in the data model. Then it sends them back to the general Alarm Queue that, in the Alarm System framework, is a Kafka data stream named Back Stage Database (BSDB).
    \item \textbf{Supervisor} is the core of the subsystem. Its task is to evaluate the inputs provided by the remote systems through the Basic Alarm Queue against the model and the defined rules (found in the Configuration Database), and ultimately generates a number of alarms, either set or cleared. The Supervisor receives also a selection of monitoring values with the aim of combining them and calculating corresponding alarms. The Supervisor manages a set of Distributed Alarm System Computing Units (DASUs), which interact with the BSDB to gather input values and generate outputs in terms of new Alarms or synthetic parameters. The DASUs output is obtained in one or more steps from the inputs by the Alarm System Computing Elements (ASCEs).
    \item \textbf{Alarm Queue} (Fig. \ref{fig_4}) serves as temporary storage both of a subset of monitoring points and actual alarm events. It also acts as a buffer to synchronize the other consumer components of the subsystem that could operate at different speeds. The Alarm Queue is the BSDB, the BASIC ALARM Queue, and the PluginQueue; it is based on Apache Kafka.
    \item \textbf{Alarm Dispatcher} is performed by a Kafka Consumer; it notifies and sends alarm data points both to the long-term alarm archive and operator GUIs.
    \item \textbf{Alarm Manager} acts as a coordinator with the task to start and stop the entire Alarm System and provides the current status of AS components.
\end{itemize}

\section{Conclusion and Future Developments}
We presented the architecture of the Monitoring, Logging, and Alarm System that monitors and logs the data needed to check and improve the operational activities of a large-scale telescope array such as ASTRI. The MLA prototype was designed and built considering the current software tools and concepts coming from Big Data and Internet of Things \citenum{10.1117/12.2560697}. The software stack is based on open-source software, thus reducing the need for unnecessary extra software development.
Future work is planned to integrate Machine Learning algorithms to perform anomaly detection and failure prediction.

\acknowledgements 
This work was conducted in the context of the ASTRI Project thanks to the support of the Italian Ministry of University and Research (MUR) as well as the Ministry for Economic Development (MISE) with funds specifically assigned to the Italian National Institute of Astrophysics (INAF). We acknowledge support from the Brazilian Funding Agency FAPESP (Grant 2013/10559-5) and from the South African Department of Science and Technology through Funding Agreement 0227/2014 for the South African Gamma-Ray Astronomy Programme. IAC is supported by the Spanish Ministry of Science and Innovation (MICIU). This work has also been partially supported by H2020-ASTERICS, a project funded by the European Commission Framework Programme Horizon 2020 Research and Innovation action under grant agreement n. 653477. The ASTRI project is becoming a reality thanks to Giovanni “Nanni” Bignami, Nicolò “Nichi” D’Amico two outstanding scientists who, in their capability of INAF Presidents, provided continuous support and invaluable guidance. While Nanni was instrumental to start the ASTRI telescope, Nichi transformed it into the Mini Array in Tenerife. Now the project is being built owing to the unfaltering support of Marco Tavani, the current INAF President. Paolo Vettolani and Filippo Zerbi, the past and current INAF Science Directors, as well as Massimo Cappi, the Coordinator of the High Energy branch of INAF, have been also very supportive to our work. We are very grateful to all of them. Nanni and Nichi, unfortunately, passed away but their vision is still guiding us. This article has gone through the internal ASTRI review process.

\bibliography{report} 
\bibliographystyle{spiebib} 

\end{document}